\documentclass[twocolumn,showpacs,preprintnumbers,amsmath,amssymb,pre]{revtex4}

\usepackage{epsfig,amsmath,amssymb,graphics,color,calc}

\renewcommand{\phi}{\varphi}
\newcommand{\taua} {\tau_{\alpha}}

\begin{document}

\title{Brambilla \textit{et al.} Reply to a Comment by J. Reinhardt \textit{et al.} on ``Probing the equilibrium dynamics of colloidal hard spheres above the mode-coupling glass transition''}

\author{G. Brambilla, D. El Masri, M. Pierno, L. Berthier, L. Cipelletti}
\affiliation{ $^2$Laboratoire des Collo\"{\i}des, Verres et
Nanomat\'{e}riaux (UMR CNRS-UM2 5587), CC26, Universit\'{e}
Montpellier 2, 34095 Montpellier Cedex 5, France }

\email{Luca.Cipelletti@univ-montp2.fr}
\date{\today}

\begin{abstract}
G. Brambilla et al. Reply to a Comment by J. Reinhardt et al.
questioning the existence of equilibrium dynamics above the critical
volume fraction of colloidal glassy hard spheres predicted by mode
coupling theory.
\end{abstract}

\pacs{64.70.pv, 05.20.Jj, 64.70.P-}
\maketitle

\textbf{Brambilla \textit{et al.} Reply:}
Reinhardt {\it et al.}~\cite{RWF}
(RWF) use mode-coupling theory (MCT)
to analyze a subset of our data~\cite{BrambillaPRL2009} and
question our claim
that dense colloidal hard spheres enter at large volume fraction
$\phi$ a dynamical regime not described by MCT. To reach this conclusion,
RWF fit intermediate scattering functions (ISFs) obtained by
light scattering to the outcome of MCT calculations
for a monodisperse system of hard spheres. By freely adjusting
the short-time diffusion coefficient $D_s$, and $w$, the parameter
fixing the relative contribution of self and collective dynamics to the
signal, they reproduce well the short-time decay of the data to a plateau. More crucially, to
reproduce also the long-time decay, RWF need to adjust, for each
experimental volume fraction $\phi$ considered, the volume fraction $\phi^{mct}$
of the corresponding theoretical curve.
Since the shape of the ISF does not change much with $\phi$,
this analysis is nearly equivalent to adjusting the typical relaxation time
$\tau_\alpha(\phi)$, which we had done more simply by fitting the data
to a stretched exponential form~\cite{BrambillaPRL2009}.

\begin{figure} [t]
\psfig{file=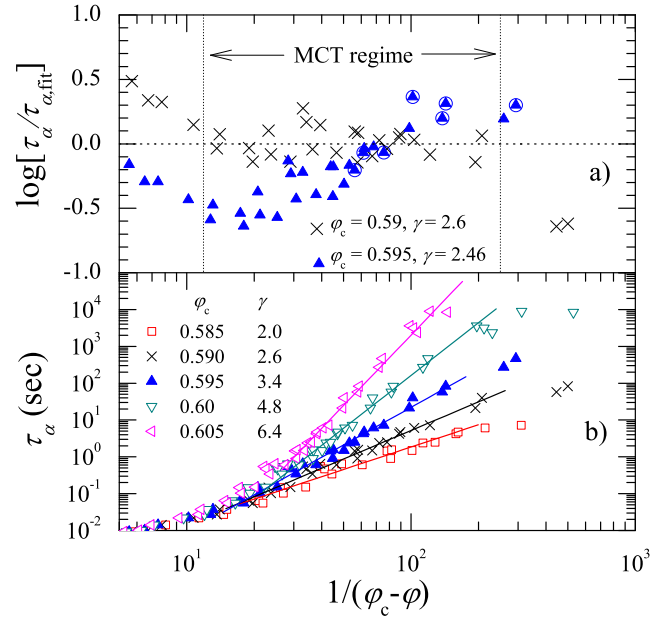,width=8.5cm,clip}
\caption{\label{fig2} (Color online)
a): Comparison of the experimental decay time of the ISF, $\tau_{\alpha}$,
to that predicted by a MCT fit, $\tau_{\alpha,\mathrm{fit}}$.
Systematic deviations are observed using RWF values, both when considering
the full set of data (triangles) or the subset
analyzed in ~\cite{RWF} (circles), while a genuine MCT regime
exists in our analysis (crosses).
b): $\taua$ vs. $(\phi_c-\phi)^{-1}$, for various choices of $\phi_c$ with
critical law fits to the data (lines), with an exponent $\gamma$
shown in labels.
Crosses correspond to $\phi_c = 0.59$, $\gamma=2.6$ as in
\cite{BrambillaPRL2009}, while solid triangles correspond
to $\phi_c = 0.595$, but with $\gamma = 3.4$, inconsistently with
\cite{RWF}.}
\end{figure}

RWF's MCT analysis differs from ours when they then estimate
the location, $\phi_c$, of what we claim is an avoided MCT transition.
If MCT predictions were an appropriate representation of our data,
the fitted $\phi^{mct}(\phi)$ should be a linear function
of $\phi$, with the critical density $\phi_c$ estimated from
$\phi^{mct}(\phi_c)=\phi^{mct}_c$, with $\phi^{mct}_c = 0.5159$.
RWF obtain $\phi_c=0.595$, although deviations from linearity are evident
in their Fig.~1b. Indeed, we find that
the value of $\phi_c$ determined according to this procedure
decreases systematically from $0.595$ to $0.590$ when the
upper limit of the fitting interval varies from $\phi=0.5908$
to $\phi=0.5852$, indicating that
the relation $\phi^{mct}(\phi)$ is not linear.
In the absence of an unambiguous criterium for selecting the
`best' $\phi_c$ from RWF analysis, it is mandatory to compare the experimental
$\taua(\phi)$ to the MCT prediction, $\taua \sim(\phi_c-\phi)^{-\gamma}$.
In Fig.~\ref{fig2}a we show that with the values $\phi_c=0.595$
and $\gamma = 2.46$ obtained by RWF, the fit deviates from the data
in a systematic manner for all $\phi$. Thus, RWF's
MCT analysis reproduces experimental ISFs but fails to
accurately determine $\phi_c$.

In Fig.~\ref{fig2}b, we show a log-log plot of $\taua$ vs.
$(\phi_c - \phi)^{-1}$,
where the MCT critical law becomes a straight line
of slope $\gamma$, thus allowing for a more stringent test
of an MCT description. We find again that an absolute
determination of $\phi_c$ is ambiguous as $\gamma$ and $\phi_c$
are correlated fitting parameter evolving from
$(\gamma=2,\phi_c=0.585)$ to $(\gamma = 6.4, \phi_c=0.605)$.
In particular, we determine $\gamma=3.4 \pm 0.1$ for the RWF preferred
value $\phi_c=0.595$, while they advocate $\gamma = 2.46$.
In our work~\cite{BrambillaPRL2009} we had used Fig.~\ref{fig2}b to determine
the best pair ($\phi_c$, $\gamma$) that fits our data.
We imposed $\gamma=2.6$, as obtained from MCT
theoretical calculations (the precise value depends of the
specific approximation used in the theory) and deduced
$\phi_c=0.59$. As shown in Fig.~\ref{fig2}a this choice opens
a genuine ``MCT regime'', which is absent in RWF's analysis.
We are then left with ISFs fully decaying to zero
for \emph{seven} samples above $\phi_c$, with significant deviations
of $\taua$ with respect to the divergence predicted by
MCT~\cite{BrambillaPRL2009,ElMasriJSTAT2009}. This
motivated us to interpret these significant deviations from MCT
predictions as the observation of a different, activated dynamical
behavior entered by colloidal hard spheres above
the divergence predicted by MCT. \\

\end{document}